\documentstyle[12pt]{article}
\def\appendix#1{
\addtocounter{section}{1}
\setcounter{equation}{0}
\renewcommand{\thesection}{\Alph{section}}
\section*{Appendix \thesection\protect\indent #1}
\addcontentsline{toc}{section}{Appendix \thesection\ \ \ #1}
}
\def\ns{\slash \! \! \! \! \nabla }

\newcommand{\del}{\frac{\tau} 2 }

\def \N {{\cal N}}
\def \ov {\over}
\def\e{{\,\rm e}\,}

\def\be{\begin{equation}}
\def\la{\label}
\def\ee{\end{equation}}
\def\bea{\begin{eqnarray}}
\def\eea{\end{eqnarray}}

\def\a{\alpha}

\def\s{\sigma}
\def\n{\nabla}

\def\D{\Delta}
\def\G{\Gamma}
\def \td {\tilde}
\def \del {\partial}
\def \na {\nabla}\def \ci{\cite}\def \foot{\footnote}
\def \ep {\epsilon}
\newcommand{\rf}[1]{(\ref{#1})}

\def\g{\gamma}
\def\d{\delta}

\def\l{\left(}
\def\r{\right)}
\def\p{\partial}

\def \bi {\bibitem}
\begin{document}
\title{\vspace{-2cm}
\hfill{\small UAHEP001} \\
\vspace{-0.5cm}
\hfill{\small OHSTPY-HEP-T-00-001}\\
\vspace{0.3cm}
Conformal anomaly of (2,0) tensor multiplet in six dimensions and  
AdS/CFT correspondence}
\vspace{-0.25cm}
\author{\large 
F. Bastianelli$^{a}$,
\mbox{} S. Frolov$^{b,}$\thanks{Also at Steklov Mathematical 
Institute, Moscow.}
\mbox{} and \mbox{} A.A. Tseytlin$^{c,}$\thanks{Also at Lebedev 
Physics Institute, Moscow and Imperial College, London.}
\mbox{}
 \\ \small  
$^a$Dipartimento di Fisica, Universit\`a di Bologna,
\vspace{-0.2cm} \mbox{} \\ \small  
V. Irnerio 46, I-40126 Bologna 
\vspace{-0.25cm} \mbox{} \\ \small 
and
\vspace{-0.25cm} \mbox{} \\ \small 
INFN, Sezione di Bologna, Italy 
\mbox{} \\ \small 
$^b$Department of Physics and Astronomy
\vspace{-0.2cm} \mbox{} \\ \small 
University of Alabama, Box 870324
\vspace{-0.2cm} \mbox{} \\ \small 
Tuscaloosa, Alabama 35487-0324, USA
\mbox{} \\ \small 
$^c$Department of Physics 
\vspace{-0.2cm} \mbox{} \\ \small 
The Ohio State University
\vspace{-0.2cm} \mbox{} \\ \small 
Columbus, OH 43210-1106, USA
\mbox{}
}
\date {}
\maketitle
\vspace{-1 cm}
\begin{abstract}
\baselineskip 11pt
We compute the conformal anomaly in the free $d=6$ superconformal
(2,0) tensor multiplet theory on generic curved background. 
Up to a trivial covariant  total-derivative term,
it is given by the sum of the type A part proportional to the  
6-d Euler density, and the type B part containing  three  
independent conformal invariants: two $CCC$  contractions 
of Weyl tensors and a $C\nabla^2C +...$ term.
Multiplied  by  the factor $4N^3$, the latter Weyl-invariant part
of the  anomaly reproduces exactly the  corresponding part of the 
conformal anomaly of  large $N$  multiple M5-brane (2,0)
theory as predicted  (hep-th/9806087) by $AdS_7$ supergravity
on the basis of AdS/CFT correspondence.
The coefficients of the type A anomaly
differ by the factor ${4 \ov 7}\times 4 N^3$, so that 
the free tensor multiplet anomaly does not vanish 
on a Ricci-flat background. The coefficient  $4N^3$ is the same
as found (hep-th/9703040) in the comparison 
of the tensor multiplet theory  and the $d=11$ 
supergravity predictions for the absorption cross-sections of 
gravitons by   M5 branes, and  in the comparison 
(hep-th/9911135) of 2- and 3-point stress tensor  correlators
of the free tensor multiplet with the $AdS_7$ supergravity
predictions.  The reason for this coincidence is that the
three Weyl-invariant terms in the anomaly are related to the 
$h^2$ and $h^3$ terms in the near flat-space 
expansion of the corresponding non-local effective action,
and thus to the 2-point and 3-point stress tensor  correlators 
in flat background. At the same time, the type A anomaly
is related to the $h^4$ term in the non-local part of the 
effective action, i.e. to a certain structure 
in the 4-point  correlation function of the stress tensors. 
It  should thus capture some  non-trivial dynamics 
of the interacting theory.
This is different from what happens in the $d=4$ SYM case where 
the type B and type A anomalies  are related to the 2-point 
and 3-point stress tensor correlators.

\end{abstract}
\vfill\eject
\baselineskip 18pt

\section{Introduction and summary}
While the  low energy dynamics of
a single M5 brane is described by the free 
$d=6, \ \N=(2,0)$ tensor multiplet, 
the low energy theory describing $N$ coincident M5 branes
 remains rather mysterious.
 One of the key predictions of the supergravity 
 description of multiple M5 branes is that the entropy \ci{kt}
 and the 2-point stress tensor  correlators \ci{gkt,gk} 
 of the large $N$ theory should scale as $N^3$. 
 Further  
 quantitative information about this interacting 
 (2,0) conformal 
 theory  
 can  be obtained using the AdS/CFT correspondence
 \cite{mal,gkp,wit}. 
 In the 
large $N$ limit 
this  leads directly to the analysis of $d=11$ supergravity 
compactified on $AdS_7\times S^4$.
In particular, spectrum of the chiral operators, some of their 2- and 
3-point functions and the structure of the anomalies have been studied
\cite{ozz,min,lei,hall,sken,harv,awata,with,AF,corr,BZ,niew,nish,imb}.

In  spite  of the lack of a useful field-theoretic 
description of the large $N$  (2,0) theory,  it is interesting 
 to 
compare its properties to those of a  $d=6$ free conformal theory 
 of a number $\sim N^3$ of  tensor 
 multiplets (after all, the free (2,0) tensor multiplet theory
 is the only $d=6$ superconformal theory  with
  the right symmetry
 properties 
 which is known explicitly).
 The idea is to try to follow  the pattern  which 
 worked in  the case of the 
  D3 brane theory 
 where certain features  of the
 strong coupling  large $N$\  $d=4, \N=4$ SYM theory 
 as described by $AdS_5 \times S^5$ supergravity 
  can be reproduced by a free theory of $N^2$    vector 
 multiplets.

In a previous paper \cite{BFT}, we have 
 found  that the 2- and 3-point correlation functions of the 
stress tensor  of (2,0) theory 
as predicted by the  $AdS_7 \times S^4$  supergravity  \cite{LT,AF}
are exactly the same as in the  theory of 
 $4 N^3$ free tensor multiplets.
 The remarkable coefficient $4N^3$ 
 is the same as found earlier in \ci{gkt} 
in the comparison 
of the  M5 brane  world volume  theory  and the $d=11$ 
supergravity expressions  for the 
absorption cross-sections of longitudinally polarized 
gravitons by  $N$ M5 branes.
This is not surprising since the ratios of the 
predictions for the 2-point stress tensor correlators
and the absorption cross sections  should be the same on
 the basis of unitarity \ci{gk,gkp}.
 That the same coefficient appears also in the ratio  
 of the 3-point correlators 
 (which in general have a   complicated 
  structure parametrized by 3 independent constants \ci{osb}) 
is quite surprising  and is likely to be a consequence of the 
extended $d=6$ supersymmetry of the theory in question.

 Here  we 
extend such a comparison to  conformal anomalies
in external $d=6$ metric. 
On the supergravity side 
of the  AdS/CFT correspondence  the conformal anomaly of 
the large $N$ M5 brane theory  was already  found 
 in \cite{sken} (see also \ci{grah}).
 Below we compute 
  the conformal anomaly  of a  free $d=6,\ \N=(2,0)$ 
  tensor multiplet  which contains  5 scalars, 2 Weyl fermions
and a chiral two-form.

In general, the trace anomaly 
in the stress tensor of a classically 
Weyl-invariant theory  in $d=2k$ dimensions has the following structure
 \ci{duff,ddi,bon,DeSc,kara}:
 $ <T> = A + B + D$, where  $A =a E_d$  is proprional to the 
 Euler density in $d$ dimensions (i.e.  is a total derivative 
 of a non-covariant expression), 
 $B=\sum_n  c_n I_n$  is a sum of independent  Weyl invariants, 
 i.e. 
 Weyl tensor contractions  with extra conformal 
 derivative operators, 
 $(C_{....})^k$,  ..., 
  $C_{....}( \nabla^{k-2} + ...) C_{....}$, 
  and $D= \n_i J^i$  is  a total derivative 
  of a covariant expression. Only type A and type B 
  anomalies are genuine (with the latter determining 
  the UV related scale anomaly), while  the type D one  
  is ambiguous (renormalization scheme dependent) as it can be 
  changed by adding  local covariant but not Weyl-invariant
  counterterms to the effective action.
  In 6 dimensions \ci{bon,kara} 
  \be
  <T> = a E_6 + (c_1 I_1 + c_2 I_2 + c_3 I_3)  + \na_i J^i  
   \ ,
   \la{ann}
    \ee
  where
  \be 
   E_6 = - \ep_6 \ep_6 R R R \ , \ \ \ 
   I_1 = C_{amnb} C^{mijn} C_i{}^{ab}{}_j \ , \ \ \ 
 I_2 =  C_{ab}{}^{mn}  C_{mn}{}^{ij} C_{ij}{}^{ab} \ , \la{inv}
  \ee
 \be
 I_3 = C_{mabc} \biggl
  (  \nabla^2 \delta^m_n+4 R^m_n -{6\over 5}R\delta^m_n
\biggr ) C^{nabc} + total~derivative
\ .  \la{inva}
\ee
Computing the conformal anomalies of the  fields
in the free  tensor multiplet and comparing  the resulting coefficients 
to the supergravity prediction \ci{sken} for the anomaly of the (2,0) 
theory  we have found that 
\be
a^{(2,0)}\   = \ {16 \ov 7}  N^3\  a^{(tens.)}\ , \ \ \ \ \ \ \ 
c^{(2,0)}_n\  = \  4 N^3 \ c^{(tens.)}_n \ . 
\la{res}
\ee
Once again,   the set  of  $4N^3$ tensor multiplets 
reproduces exactly   the type B or  scale 
anomaly of the (2,0)  theory!
 However, the coefficients 
of the type A   anomaly then differ. 

The ratio $4N^3$ of the $c_n$ coefficients 
is, in fact,  in direct correspondence  with  the 
result for the ratio of the 2- and 3-point correlators 
of the stress tensor found in \cite{BFT}.
At the same time, the coefficient $a$ of the  Euler density term 
in the anomaly turns out to be related 
to a coefficient of a certain structure 
in the {\it 4-point } correlation function of the stress tensor
 and  should thus reflect 
some of the non-trivial dynamics 
of the interacting theory.\foot{Note  that the type A anomaly 
 coefficient $a$ 
(in any dimension) 
plays a special role from the point of view 
of the supergravity analysis \ci{imb}.}

To appreciate this novel feature of the $d=6$ theory it is useful 
to compare the above  results 
with what happens in 
the $d=4$ case  --  the ${\cal N}=4$ 
super Yang--Mills theory.
In $d=4$ the type B anomaly contains
 just one independent term proportional to 
the square of the Weyl tensor 
whose coefficient is  directly related to  the one in 
 the 2-point function $<TT>$ 
of the stress tensor.
The type A (Euler) anomaly 
is instead  related  to the 3-point correlation function 
 $<TTT>$.
Thus known non-renormalization theorems for the  2- and 3-point
functions of the stress tensor multiplet 
guarantee that 
the  trace anomaly of $N^2$ free $\N=4$  vector multiplets
should reproduce that of the full interacting non-abelian theory
(see \ci{nonr,gk}).

In $d=6$ the coefficient $a$ is related to the
 4-point function and thus
there is no reason to expect that it should
 not be renormalized.\foot{If the $d=6$ free and 
 interacting CFT's 
 discussed  above could 
be linked by a renormalization group flow preserving maximal
supersymmetry,
then our results would suggest that only the 
coefficient $a$ of  type A anomaly can flow. 
However, it is difficult to see how such 
 picture could be realized
since  the interacting theory at large $N$ does not have  
suitable scalar operators of dimensions $\Delta \leq 6$ which
 could be used to deform the
theory (the only candidates are charged under 
the R symmetry and would break
maximal supersymmetry).
Similarly, the cohomological analysis of 
\cite{hano} indicates that a
theory containing a free chiral
 two-form field  cannot be continuously deformed
in a non-trivial manner.}
It would be interesting  to see if the $R^4$ correction 
to the $d=11$ supergravity action generates  an order $N$ 
correction to the coefficient $a$ in the supergravity expression 
for the conformal anomaly, 
like it does in the entropy of multiple M5 branes \ci{GKTM}.


The above mentioned correspondence  between 
particular terms in the conformal anomaly 
and correlation functions  of stress tensor 
on flat background can be understood  by 
studing the relation between the type A and type B 
 conformal anomalies  and   corresponding terms in the
  effective action following \ci{Deser,DeSc,kara,anse}.
  In a  general even dimension $d=2k$  the Weyl-invariant 
  terms  in type B  part of the conformal anomaly
  $(C_{....})^k$,  ..., 
  $C_{....}( \nabla^{k-2} + ...) C_{....}$
   can be 
  obtained by the  Weyl variation 
   from the non-local scale-dependent  terms in the 
  effective action like 
  $\int (C_{....})^{k-1} \ln( \mu^{-d} \td \Delta_d) C_{....}, ...$,
  $\int C_{....}\td  \Delta_d \ln( \mu^{-d} \td
   \Delta_d) C_{....}$.
  Here  $\td \Delta_d = \nabla^d + ...$ is an appropriate 
  `Weyl-covariant' operator acting on Weyl tensor \ci{Deser}.
  Expanded near flat space, $g_{mn} = \d_{mn} + h_{mn}$, 
  these terms  start with $h^k, ..., h^2$, respectively,  
  i.e. correspond  to  particular 
  structures in the  $k-, ..., 2-$ point correlators of the 
  stress tensor in flat background, respectively. 
  
  In the case of $d=4$, i.e. $k=2$, the coefficient 
  of the type B anomaly ($C^2$)  
  is thus correlated with the coefficient 
  in the 2-point  function $<TT>$. 
  In the case of $d=6$, i.e. $k=3$, 
  the coefficient $c_3$ of the  $I_3$ term in \rf{ann}
   corresponds 
  to the one in the  2-point function, while 
  the two other  coefficients $c_1,\ c_2$ 
  should be directly related to the two remaining independent 
  coefficients in the generic $d=6$ CFT  
  correlator $<TTT>$.\foot{In  a  generic $d=6$  CFT 
   the 3-point  stress tensor 
  correlator depends 
  on 3 arbitrary parameters but one combination  of them 
  is related by Ward identity to the coefficient 
  in the 2-point function \ci{osb}.}
  
As for the  type A anomaly, 
the corresponding term in the effective action 
can be constructed  by integrating the conformal anomaly 
like it   was done in 2 \ci{pol} and 4 \ci{rig,ft} dimensions.
One can introduce the modified Euler density \ci{anse}
by combining the type A anomaly with a particular type D anomaly, 
$\td E_d = E_d + \na_i \td J^i$, 
where $\td J^i$ is a  covariant expression  such that the Weyl variation 
$(\delta g_{mn} = \phi g_{mn}$) 
of  $\td E_d$ is proportional to $\Delta_d \phi$  where
$\Delta_d = \nabla^d + ... $ is the Weyl-invariant  operator
acting on scalars  (see, e.g., 
 \ci{erd} and refs. there).
 Then the corresponding term in the effective action 
 is \ci{Deser} \
  $\int \td E_d { 1 \ov \Delta_d}  \td E_d$.
 Expanding this term near flat space  and discarding 
 local terms (which correspond to contact terms in the 
 stress tensor correlators) it is possible to argue that
  the leading  non-local structure
   with single 
  $\del^{-2}$ pole  contains $k+1$  graviton factors.
  Thus the type A anomaly term is related  
   to the ${d \ov 2} + 1$ point 
  correlator of stress tensors, i.e. to 3-point correlator 
  in $d=4$, 
  4-point correlator  in $d=6$, etc.\foot{For example, in $d=4$  \ci{Deser}\ 
  $\Delta_4 \sim \del^4 + \del R_0 \del + ...$, $R_0 \sim \del \del h$, 
  $\td E_4 \sim  E_4 +  \del^2 R_0  $, so that 
  $\int  \td E_4 { 1 \ov \Delta_4}  \td E_4 \sim \int R_0 \del^{-2}
   \del R_0 \del \del^2 R_0 + ...$\ .
   In $d=6$   $\td E_6$ has the structure 
   (see \ci{anse})
     $  E_6 + R_0 \del^2 R_0  + \del^4 R_0 $  and one finds that 
     the first  non-local term in  
       $\int  \td E_6 { 1 \ov \Delta_6}  \td E_6 $ is 
       of order $h^4$.}

Coming back to \rf{res}, 
the disagreement of the total expressions
 for the conformal anomalies of the free tensor multiplet and
 interacting (2,0) CFT 
can be easily seen  using the results which already 
existed in the literature. Choosing  a  Ricci-flat 
\ $d=6$ background one  finds that (2,0) theory anomaly 
found in  \ci{sken} vanishes,  but the combined  anomaly 
of the fields in the tensor multiplet is {\it non-zero}.
For $R_{mn}=0$ the  $d=6$ anomaly \rf{ann}
depends (modulo a covariant total derivative term)
only on 2 coefficients and can be written in 
the form \ci{FT}:\foot{In the notation of \ci{FT}
$E_6 =  -32 E$  with the Euler number being $\chi = { 1 \ov (4 \pi)^3}
\int d^6 x \sqrt g  \ { 2 \ov 3} E$ and the invariants 
 called 
$I_1,I_2$ in \ci{FT} are $A_{16},A_{17}$, or, for $R_{mn}=0$, 
$I_2, - I_1$ in the present paper.}
\ $<T> = { 1 \ov (4 \pi)^3} (s_1  E_6   + s_2  I_2)  .$
 The coefficients $s_1,s_2$ are linear combinations
 of $a,c_1,c_2,c_3$ in \rf{ann}.
The coefficient $s_2$ is proportional to the graded number of degrees of
freedom  and thus vanishes ($5+3-2 \times 4=0$) 
for the tensor multiplet. Computing the coefficient
$s_1$ by combining the known results for a $d=6$ scalar 
\ci{sakai,Gil,dowk},
spinor \ci{crit} and  the antisymmetric tensor \ci{FT}
(using that 
the conformal anomaly of the chiral 2-form field is half 
of the anomaly  of a non-chiral 2-form)
one finds that (see eqs. (4.21),(4.39) in \ci{FT})
$s_1= 
   - { 1 \ov 32}\times  {
 1 \ov 12}$.\foot{It is easy to check that this anomaly cannot be cancelled
by adding, e.g., 
  a non-dynamical 5-form field
 which carries no degrees of freedom ($s_2=0$)
 but does produce a non-trivial  conformal anomaly \ci{duvn,FT}
 $s_1= 
  - { 1 \ov 32}\times 2$.
   Duality rotation of scalars into 4-form fields 
 also does not help ($s_1^{(4-form)} - s_1^{(0-form)}$
 = $ { 1 \ov 32}\times { 4 \ov 3}$), and, in any case,  
 the duality transformation is 
   not consistent with conformal invariance.}
 Thus, in contrast to the $d=4$  case  where the conformal anomaly 
 of the $\N=4$ vector multiplet vanishes 
 on the Ricci flat background, 
 there does not seem to exist a free $d=6$  conformal
 matter  theory\foot{Note, however,  that the  combined Seeley
 coefficient $b_6$  of the 
fields of the $d=11$ supergravity or its reduction 
to 6 dimensions does  vanish \ci{FT}.}
  which shares the same property.\foot{Possible importance of theories  in which the coefficients 
 $a$ and $c_i$ are related in such a way that the anomaly vanishes for
 $R_{mn}=0$ was advocated in  \ci{ansel}.}

Our aim below in Section 2 will be to find 
  the 
 scalar, spinor and 2-form anomalies
  without assuming
   $R_{mn}=0$ 
    and thus to be able 
  to compare the tensor multiplet anomaly 
   to the supergravity prediction  \ci{sken}
   for the anomaly of the (2,0) theory.
 Our starting point will be the general expression \ci{Gil}
 for the  corresponding Seeley coefficient $b_6$ (sometimes 
 called also  $a_3$)
 of the second  order Laplacian.
 While the scalar field anomaly
  was computed in the past \ci{Gil,toms,PR,ichi}
  (though was not correctly put   into the 
  required conformal basis form), 
 our explicit expressions for the spinor 
 and the 2-form anomalies are new.
 
 Appendix A  contains conventions and definitions of 
 the basic  curvature invariants in $d=6$.
 In Appendix B we present
 the full expression for the 
 Seeley coefficient $b_6$ taken from \ci{Gil}
 but written in a slightly different form.
 In Appendix C we give  the results for the $b_6$ coefficients
 for the 1-form and 2-form fields in a space of 
 general dimension $d$.

\section{Conformal anomaly of $d=6$ \  (2,0) tensor multiplet}
The low-energy effective theory of a single 
 M5-brane is described by 
the (2,0) tensor multiplet 
consisting of 5 scalars $X^a$, an 
antisymmetric tensor $B_{ij}$ with (anti)selfdual strength
and 2 Weyl fermions $\psi^I_L$. 
It is sufficient for the purposes of computing the conformal anomaly 
to consider 
 the non-chiral (2,2)  conformal model described 
by the following action 
\bea
S=\int\, d^6x\,\sqrt{g} \biggl( -\frac{1}{12}H_{ijk}^2-
\frac12\n_i X^\a\n^i X^\a
-\frac{1}{10} X^\a X^\a R 
+i\bar{\psi}^I\g^m\n_m\psi^I\biggr) , 
\la{lagr}\eea
where $H_{ijk}=\p_i B_{jk} +\p_j B_{ki}+\p_k B_{ij}$, 
\ \ $i,j,k=1,...,6$, $\a=1,...,10$, and $I=1,2$.
The trace anomaly of the (2,0) tensor multiplet is 
then equal to 1/2 of the trace anomaly of the (2,2) multiplet. 
Indeed, we can consistently disregard the gravitational anomalies
of the (2,0) multiplet
related to the imaginary part of the chiral 2-form and Weyl 
spinor  determinants
and focus only on their real part leading to the trace anomalies
(see \cite{AGW,BFT}).\footnote{While 
 the definition of the
partition function 
of a  chiral p-form theory  on generic manifolds 
is subtle (see \ci{witte,hen}) 
the coefficients in the local trace anomaly  are universal 
(cannot depend
on  details of space-time topology) and can be determined, e.g., 
by a Feynman  diagram calculation near flat space-time.}
The  anomaly of the (2,2) theory  is given by the sum of
trace anomalies of one non-chiral 2-form, 
10 conformal scalar fields and 2 Dirac fermions
which we shall compute separately below.

\subsection{Conformal anomaly as  Seeley-DeWitt coefficient}

We  begin by recalling the relation of free-theory 
conformal anomaly to 
the Seeley-DeWitt coefficients. Consider a one-loop approximation
to a model of a bosonic field $\phi$ taking values in a smooth 
vector bundle $V$ over 
a compact smooth Riemannian manifold $M$ of dimension $d$. The partition 
function and the effective action of the model are given by
\bea
Z&=&\int\, d\phi \, \e^{-\frac12 \int_M\,\phi\D\phi},\qquad 
\G = \frac12 {\rm log\, det}\ \D\ , 
\nonumber\\
\D &=&-\n^2 -  E\ , \la{oper}
\eea
where $\n$ is a covariant derivative  on $V$ and the matrix function 
$E$ is an endomorphism of $V$.
It is well-known that the trace anomaly is related to the
logarithmically divergent part of the effective action. 
Using the Seeley-DeWitt asymptotic expansion \cite{DW, See}
\bea
({\rm Tr} \e^{-s\D})_{s\to 0} \sim \sum_{p=0}^d\, s^{\frac12 (p-d)}
\int_M\, d^dx\ \sqrt{g} \ b_p
\nonumber
\eea
we get the logarithmically divergent term  in the effective action
$$
\G _\infty =-\frac12 {\rm log}\frac{L^2}{\mu^2}\,
\int_M\, d^dx\ \sqrt{g}\  b_d\ ,$$
where $L\to \infty$ is an UV cut-off. The trace anomaly of the stress 
tensor is then equal to $b_d$:
\be 
\langle T_m{}^m(x)\rangle = b_d(x) \  .
\ee
Thus to find the conformal anomaly of the $d=6$  tensor multiplet we
need to know the coefficients $b_6$ for  various second  order Laplace
operators corresponding to the fields in \rf{lagr}. 

The coefficient $b_6$ was explicitly 
computed for any operator of the form (\ref{oper})
in \cite{Gil}, and can be written as follows
\bea
b_6(\D )&=&\frac{1}{(4\pi )^3 7!}{\rm tr}_V \biggl[
18A_1+17A_2-2A_3-4A_4+9A_5\cr
&+&28A_6-8A_7
+24A_8+12A_9+
\frac{35}{9}A_{10}-\frac{14}{3}A_{11}+\frac{14}{3}A_{12}\cr
&-&
\frac{208}{9}A_{13}+\frac{64}{3}A_{14}-\frac{16}{3}A_{15}+
\frac{44}{9}A_{16}+\frac{80}{9}A_{17}\cr
&+&
14\biggl( 8V_1+2V_2+12V_3-12V_4+6V_5-4V_6+5V_7\cr
&+&6V_8+60V_9
+30V_{10}+60V_{11}+30V_{12}+10V_{13}+4V_{14}\cr
&+&12V_{15}+
30V_{16}+12V_{17}+5V_{18}-2V_{19}+2V_{20}\biggr)\biggr] ,
\la{b6}
\eea
where the invariants $A_s$ and $ V_p$ (depending on the metric tensor, 
the connection curvature tensor and the endomorphism $E$) are 
listed in the Appendix $A$.
An explicit   expression for $b_6$ can be found  in Appendix B.

As already discussed in the Introduction, the conformal anomaly
in a classically Weyl-invariant $d=6$ theory,   or 
 the coefficient $b_6$ for a conformally invariant 
kinetic operator,  must  have  the form \cite{bon, DeSc, kara}
\bea
b_6= a E_6+c_1I_1+c_2I_2+c_3I_3+\n_iJ^i\ .
\la{invform}
\eea
Here the first term is 
the type A anomaly proportional to the Euler density  polynomial
\bea
E_{6} &=& -\epsilon_{m_1n_1m_2n_2m_3n_3}
\epsilon^{a_1b_1a_2b_2a_3b_3}
R_{a_1b_1}^{m_1n_1} R_{a_2b_2}^{m_2n_2} R_{a_3b_3}^{m_3n_3} 
\cr
&=&  -8 A_{10} +96A_{11} -24A_{12} -128A_{13} -192A_{14} 
+192 A_{15} -32A_{16}+64A_{17}\ , 
\la{G6}
\eea
while the next three terms represent the type B anomalies
that are combinations of the following
Weyl invariants 
\bea
I_1 &=& C_{amnb} C^{mijn} C_i{}^{ab}{}_j \cr
&=& \frac{19}{800} A_{10}-\frac{57}{160} A_{11} +
\frac{3}{40} A_{12} +\frac{7}{16} A_{13} +\frac{9}{8} A_{14} 
-\frac{3}{4} A_{15} -A_{17}\ , 
\la{I1}\\
I_2&=& C_{ab}{}^{mn}  C_{mn}{}^{ij} C_{ij}{}^{ab} \cr
&=& \frac{9}{200}  A_{10} -\frac{27}{40} A_{11} 
+\frac{3}{10} A_{12} +\frac{5}{4} A_{13}+\frac{3}{2} A_{14} 
-3A_{15} +A_{16}\ , 
\la{I2}\\
I_3 &=& C_{mabc} \biggl (  \nabla^2 \delta^m_n+4 R^m_n -{6\over 5}R\delta^m_n
\biggr ) C^{nabc} +\n_i {\cal J}^i\la{I3}\\
&=& -\frac{11}{50} A_{10} +\frac{27}{10} A_{11} 
-\frac{6}{5} A_{12} -3A_{13} -4A_{14} 
+4A_{15}+\frac{1}{10} A_{6}-A_{7}+A_{9}+\n_i {\cal J}^i\ , 
\nonumber
\eea
where
\be 
C_{abcd} = R_{abcd} -{1\over 4}( g_{ac} R_{bd} + g_{bd}R_{ac}
-g_{ad}R_{bc} -g_{bc}R_{ad}) +{1\over 20}(g_{ac}g_{bd} - g_{ad}g_{bc}) R 
\ee
is the Weyl tensor in 6 dimensions,  
$$\n_i {\cal J}^i=5C_5-8C_7=3A_{3}-6A_{4}+3A_5+\frac12 A_6-5A_7+5A_9
+2A_{13}-2A_{14}-4A_{15}+2A_{16}+8A_{17},$$
and the invariants $C_k$ are defined in Appendix A.
The invariant $I_3$ was defined  up to the total derivative term 
$\n_i {\cal J}^i$ in \cite{anse}
  (similar  invariants in  \cite{bon, DeSc, erd}
are linear combinations of $I_3$ with  the other invariants), 
and is related to the invariant $\Omega_6$ used in \cite{PR} as 
$I_3=3\Omega_6+16I_1-4I_2$. 
Finally, the last term in eq. (\ref{invform})
is a total derivative of a covariant expression 
which can be cancelled by the Weyl  variation
of a finite local covariant counterterm.
 Thus, only the coefficients of the first four terms in \rf{invform}
have unambiguous  (scheme-independent) 
meaning and will be of our main 
interest below.

\subsection{Conformal anomaly of a scalar field} 

In the simplest case of a $d=6$ conformal scalar field 
the Laplace operator $\D$ is given by
\bea
\D_S = - \nabla^2  + {1\over 5} R \ ,
\la{DS}
\eea
where
the connection in $\n$ is trivial ($F_{ij}=0$), and 
the endomorphism $E$
is 
$$E =-{1\over 5} R\ .$$
A straightforward calculation based on  \rf{b6}  gives  the trace anomaly of 
the conformal
scalar as 
\bea
{\cal A}_S&=&\langle T^S_m{}^m(x)\rangle = b_6^S(x)=\cr
&=&\frac{1}{(4\pi )^3\, 7!}\biggl( \frac{6}{5}A_{1}+\frac{1}{5}A_{2}
-2A_{3}-4A_{4}+9A_{5}\cr
&-&8A_{7}+\frac{8}{5}A_{8}+12A_{9}
-\frac{7}{225}A_{10}+\frac{14}{15}A_{11}-\frac{14}{15}A_{12}
-\frac{32}{45}A_{13}\cr
&-&\frac{16}{15}A_{14}-\frac{16}{3}A_{15}
+\frac{44}{9}A_{16}+\frac{80}{9}A_{17}\biggr)
\la{b6s} \ .
\eea
This formula can be rewritten in the form (\ref{invform}) by using 
the identities  from Appendix A:
\bea
{\cal A}_S
=\frac{1}{(4\pi )^3\, 7!}\biggl( -\frac{5}{72}E_{6}-\frac{28}{3}I_{1}
+\frac{5}{3}I_{2}+2I_{3}+\n_iJ^i\biggr) \ ,
\la{b6ss}
\eea
where
$$ \n_iJ^i=\frac{6}{5}C_{1}-\frac{2}{5}C_{2}
+4C_{3}+\frac{12}{5}C_{4}
+\frac{17}{5}C_{6}+12C_{7}\ .
$$
Note that our  expression \rf{b6ss}
 differs from the one derived in
\cite{PR}.

\subsection{Conformal anomaly of a Dirac fermion} 

The square of the Dirac 
operator gives  the following second 
order differential operator $\D_F$ 
\bea
\D_F =  - (\ns)^2 =- \nabla^2  + {1\over 4} R \cdot 1\  .
\la{DF}
\eea
The  connection in $\n$ in this case is nontrivial  with 
$$F_{ij}=\frac14 R_{ijab}\g^{ab}$$ 
and the endomorphism $E$ 
is 
$$E =-{1\over 4} R\cdot 1\ .$$
The calculation  of the corresponding $b_6$ coefficient 
\rf{b6}  gives the following expression 
for the trace anomaly of a Dirac fermion
(we account for the Fermi statistics  by reversing the sign
of $b_6$)
\bea
{\cal A}_F&=&\langle T^F_m{}^m(x)\rangle = -b_6^F(x)=\cr
&=&-\frac{8}{(4\pi )^3\, 7!}\biggl( -3A_{1}+\frac{5}{4}A_{2}
-9A_{3}+3A_{4}-5A_{5}+\frac72 A_6\cr
&-&8A_{7}-4A_{8}-9A_{9}
-\frac{35}{72}A_{10}+\frac{7}{3}A_{11}+\frac{49}{24}A_{12}
+\frac{44}{9}A_{13}\cr
&-&\frac{20}{3}A_{14}+\frac{5}{3}A_{15}
-\frac{101}{18}A_{16}-\frac{109}{9}A_{17}\biggr)\ . 
\la{b6f}
\eea
By using the identities  from Appendix A
we can rewrite this  in the form (\ref{invform}) 
\bea
{\cal A}_F
=\frac{1}{(4\pi )^3\, 7!}\biggl(- \frac{191}{72}E_{6}-\frac{896}{3}I_{1}
-32I_{2}+40I_{3}+\n_iJ^i\biggr)\  ,
\la{b6ff}
\eea
where
$$\n_iJ^i= 24C_{1}-\frac{148}{15}C_{2}
+136C_{3}+48C_{4}-168C_{5}
+96C_{6}+352C_{7}\ .
$$

\subsection{Conformal anomaly of a  2-form field } 

To find the conformal anomaly of an antisymmetric tensor field
 we 
use a covariant gauge fixing with the standard triangle-like
ghost structure \ci{antis}.
 This leads to the following representation for the 
partition function
\bea
Z_{(2)}=({\rm det}\D^{(2)})^{-\frac12}
\ {\rm det}\D^{(1)} \ ({\rm det}\D^{(0)})^{-\frac32}\ ,
\la{Z2}
\eea
where the Hodge-DeRham operators $\D^{(p)}$ are defined as 
\bea
({\D}^{(2)})_{mn}^{ab} &=& -  \nabla^2 \d_{mn}^{ab} 
+2 R_{[m}^{[a} \d_{n]}^{b]}
- R_{mn}{}^{ab}\ , 
  \cr
({\D}^{(1)})_m^n &=& - \nabla^2  \d_m^n  + R_m^n \ , \cr
{\D}^{(0)} &=& - \nabla^2\ . \nonumber
\eea
As follows  from (\ref{Z2}),  
the conformal anomaly of a 2-form field $B_{mn}$ is given by
\bea
{\cal A}_B=b_6^{(2)}-2b_6^{(1)}+3b_6^{(0)}\ ,
\la{b6B}
\eea
where $b_6^{(p)}$ are the Seeley-DeWitt coefficients of the 
operators $\D^{(p)}$.

The coefficient $b_6^{(0)}$ is obtained from (\ref{b6}) by 
dropping out all the invariants 
$V_p$ (in this case of ${\D}^{(0)}$  the connection and
 $E$  are
trivial)
\bea
b_6^{(0)}&=&\frac{1}{(4\pi )^3 7!}\biggl(
18A_1+17A_2-2A_3-4A_4+9A_5\cr
&+&28A_6-8A_7
+24A_8+12A_9+
\frac{35}{9}A_{10}-\frac{14}{3}A_{11}+\frac{14}{3}A_{12}\cr
&-&
\frac{208}{9}A_{13}+\frac{64}{3}A_{14}-\frac{16}{3}A_{15}+
\frac{44}{9}A_{16}+\frac{80}{9}A_{17}
\biggr)\  .
\la{b60}
\eea
The coefficient $b_6^{(1)}$ of the Hodge-DeRham operator $\D^{(1)}$ acting 
on 1-forms is found by taking into account that the connection
is defined by the Christoffel symbols so that  
$$\l F_{ij}\r_{a}{}^{b}=R_{ija}{}^{b}\ ,$$
while the endomorphism $E$ is 
$$E_{a}{}^{b}=- R_a{}^b\ .$$
Computing the invariants $V_p$ and expressing them in terms  of 
 $A_s$, we get
 for $d=6$ \footnote{We present the coefficients 
$b_6^{(1)}$ and $b_6^{(2)}$ for generic dimension $d$ of the manifold
in Appendix C.}
\bea
b_6^{(1)}(x)
&=&\frac{1}{(4\pi )^3\, 7!}\biggl( 24A_{1}-66A_{2}
+352A_{3}+32A_{4}-58A_{5}-140A_6\cr
&+&792A_{7}+32A_{8}-96A_{9}
-\frac{140}{3}A_{10}+420A_{11}-70A_{12}
-\frac{2600}{3}A_{13}\cr
&+&16A_{14}+444A_{15}
-\frac{164}{3}A_{16}-\frac{344}{3}A_{17}\biggr) .
\la{b61}
\eea
To compute the coefficient $b_6^{(2)}$ of the 
operator $\D^{(2)}$ acting on 2-forms, we note  that the curvature tensor of
the connection and the endomorphism here  are
\bea
\l F_{ij}\r_{ab}^{cd}&=&2R_{ij[a}{}^{[c}\d_{b]}^{d]}\ ,  \cr
E_{ab}^{cd}&=&-2 R_{[a}^{[c} \d_{b]}^{d]}
+ R_{ab}{}^{cd}\ . 
\nonumber
\eea
By using the  formulas for the traces  from Appendix C, 
we find in  $d=6$
\bea
b_6^{(2)}(x)
&=&\frac{1}{(4\pi )^3\, 7!}\biggl( -66A_{1}+3A_{2}
-254A_{3}+164A_{4}+107A_{5}+28A_6\cr
&-&120A_{7}-88A_{8}+348A_{9}
+\frac{595}{3}A_{10}-2478A_{11}+518A_{12}
+\frac{10384}{3}A_{13}\cr
&+&4912A_{14}-4896A_{15}
+\frac{2992}{3}A_{16}-\frac{1616}{3}A_{17}\biggr)\ .
\la{b62}
\eea
Then from  (\ref{b6B}) we obtain the conformal anomaly of
the 2-form field $B_{ij}$ 
\bea
{\cal A}_B&=&\langle T^B_m{}^m(x)\rangle \cr
&=&\frac{1}{(4\pi )^3\, 7!}\biggl( -60A_{1}+186A_{2}
-964A_{3}+88A_{4}+250A_{5}+392 A_6\cr
&-&1728A_{7}-80A_{8}+576A_{9}
+\frac{910}{3}A_{10}-3332A_{11}+672A_{12}
+\frac{15376}{3}A_{13}\cr
&+&4944A_{14}-5800A_{15}
+\frac{3364}{3}A_{16}-\frac{848}{3}A_{17}\biggr)\ . 
\la{b6BB}
\eea
One can easily check that on a Ricci flat manifold
 the anomaly coincides
with the one found in \cite{FT}.

The identities  from Appendix A allow 
to represent (\ref{b6BB}) in the required  
form (\ref{invform}) 
\bea
{\cal A}_B
=\frac{1}{(4\pi )^3\, 7!}\biggl( -\frac{221}{4}E_{6}-\frac{8008}{3}I_{1}
-\frac{2378}{3}I_{2}+180I_{3}+\n_iJ^i\biggr) \ ,
\la{b6BBf}
\eea
where
$$\n_iJ^i=-60C_{1}+\frac{2036}{15}C_{2}
-1152C_{3}-120C_{4}-504C_{5}
-646C_{6}+856C_{7}\ .
$$
\subsection{Conformal anomaly of the free (2,0) tensor multiplet} 
Finally, all is prepared  
to write down the expression for the 
conformal anomaly  of the chiral (2,0) tensor multiplet
\bea
{\cal A}_{tens.}&=&
\frac12\l {\cal A}_{B}+10{\cal A}_{S}+ 2{\cal A}_{F}\r \cr
&=&\frac{1}{(4\pi )^3\, 7!}\biggl( 84A_{2}
-420A_{3}+210A_{5}+168 A_6
-840A_{7}+420A_{9}
+\frac{777}{5}A_{10}\cr
&-&1680A_{11}+315A_{12}
+2520A_{13}
+2520A_{14}-2940A_{15}
+630A_{16}\biggr)\ ,
\la{b6TM}
\eea
or,  in the form (\ref{invform}), 
\bea
{\cal A}_{tens.}
=\frac{1}{(4\pi )^3\, 7!}
\biggl( -\frac{245}{8}E_{6}-1680I_{1}
-420 I_{2} + 140 I_{3}+\n_iJ^i\biggr) ,
\la{b6TMf}
\eea
where
$$\n_iJ^i= 56C_{2} - 420C_{3} -420C_{5} - 210C_{6} +840C_{7}.
$$
It is easy to see using the identities in Appendix A
that for $R_{mn}=0$  this expression  agrees with the expression 
following from \ci{FT} which was already
mentioned in the Introduction, i.e. 
${\cal A}_{tens.}
=- \frac{1}{(4\pi )^3 \times 32 \times 12}  E_{6}$, 
up to a covariant
 total derivative term 
($\sim C_5 = { 1 \ov 2} \nabla^2 (C^2_{mnkl})$).\foot{Note 
that for
$R_{mn}=0$  and ignoring the total derivative term
one has the following relations 
$I_3= -I_2 + 4 I_1,$ \ $E_6 = - 32 I_2 - 64 I_1$. }

Let us now compare the  result \rf{b6TMf}
with  the  conformal anomaly of the interacting (2,0)  
theory describing large 
number $N$  of coincident M5 branes  as  predicted 
on the basis of AdS/CFT correspondence  
  in \ci{sken}.
 In terms of the invariants we are using here 
the expression  obtained in   \cite{sken}  takes the form 
(note that it vanishes for $R_{mn}=0$ as it should) 
\bea
{\cal A}_{(2,0)}
=\frac{4N^3}{(4\pi )^3\, 7!}
\biggl( -\frac{35}{2}E_{6}
-1680 I_{1} - 420I_{2}+ 140I_{3}+\n_iJ^i\biggr) \ ,
\la{b6HS}
\eea
where
$$\n_iJ^i=420C_{3}-504C_{4}-840C_{5}-84C_{6}+1680C_{7}.
$$
Comparing   \rf{b6TMf}  and  \rf{b6HS}
we conclude  that up to the common factor  $4N^3$ 
only the coefficient in front of the Euler polynomial
is different 
(the difference in coefficients of 
 total derivative terms is not important since
  they are scheme-dependent). 
The  interpretation of this result was already discussed 
in the Introduction.

\vfill \eject

\section*{Appendix A: Conventions,  invariants  and identities}
We use the following conventions for the  curvature tensors:
$$\biggl [ \nabla_a, \nabla_b] V^c = R_{ab}{}^c{}_d V^d \ , \ \ \ 
R_{ab}= R_{ca}{}^c{}_b\  ,\ \ \ \  R= R^a_a \ , \ \ \ 
[ \nabla_a, \nabla_b]\phi = F_{ab}\phi \ .$$ 
The basis of  metric invariants is\footnote{We use the same 
notation as in \cite{PR}. Note, however, that there are a number of
misprints in that paper.}
\bea
&&A_1= \nabla^4 R,\quad A_2=(\nabla_a R)^2,\quad 
A_3=(\nabla_a R_{mn})^2 ,~ A_4=\nabla_a R_{bm} \nabla^b R^{am},
~ A_5= (\nabla_a R_{mnij})^2 , \cr
&&A_6= R \nabla^2 R,\quad
A_7= R_{ab} \nabla^2 R^{ab},\quad 
A_8= R_{ab} \n_m\n^b R^{am},\quad
A_9= R_{abmn} \nabla^2 R^{abmn},~
A_{10}= R^3 ,\cr
&&A_{11}= R R_{ab}^2 ,\quad
A_{12}= R R_{abmn}^2 ,\quad
A_{13}= R_a{}^m R_m{}^i R_i{}^a ,\quad
A_{14}= R_{ab} R_{mn} R^{ambn},\cr
&&A_{15}= R_{ab} R^{amnl} R^b{}_{mnl},\quad
A_{16}= R_{ab}{}^{mn} R_{mn}{}^{ij}R_{ij}{}^{ab},\quad
A_{17}= R_{ambn} R^{aibj} R^{m}{}_i{}^n{}_j\ . 
\nonumber
\eea
Another convenient basis of the metric invariants is obtained 
by replacing
the Ricci tensor $R_{ij}$ by its traceless part 
$$B_{ij}=R_{ij}-\frac{1 }{d}Rg_{ij} \ , $$ and the Riemann tensor -- 
by the Weyl  tensor 
$$C_{ijkl}=R_{ijkl}-\frac{1}{d-2}(g_{jl}B_{ik}-g_{jk}B_{il}+
g_{ik}B_{jl}-g_{il}B_{jk})-
\frac{R}{d(d-1)}(g_{jl}g_{ik}-g_{jk}g_{il})\ .$$
Then we get the following 17 invariants $B_s$ 
\bea
&&B_1= \nabla^4 R,\quad B_2=(\nabla_a R)^2,\quad 
B_3=(\nabla_a B_{mn})^2 ,~ B_4=\nabla_a B_{bm} \nabla^b B^{am},
~ B_5= (\nabla_a C_{mnij})^2 , \cr
&&B_6= R \nabla^2 R,\quad
B_7= B_{ab} \nabla^2 B^{ab},\quad 
B_8= B_{ab} \n_m\n^b B^{am},\quad
B_9= C_{abmn} \nabla^2 C^{abmn},~
B_{10}= R^3 ,\cr
&&B_{11}= R B_{ab}^2 ,\quad
B_{12}= R C_{abmn}^2 ,\quad
B_{13}= B_a{}^m B_m{}^i B_i{}^a ,\quad
B_{14}= B_{ab} B_{mn} C^{ambn},\cr
&&B_{15}= B_{ab} C^{amnl} C^b{}_{mnl},\quad
B_{16}= C_{ab}{}^{mn} C_{mn}{}^{ij}C_{ij}{}^{ab},\quad
B_{17}= C_{ambn} C^{aibj} C^{m}{}_i{}^n{}_j\ . 
\nonumber
\eea
The invariants $A_s$ are related to $B_s$ as follows
\bea
&&A_1=B_1,\quad A_2=B_2,\quad A_3=B_3+\frac{1}{d} B_2,\quad
A_4=B_4+\frac{d-1}{d^2}B_2\cr
&&A_5=B_5+\frac{4}{d-2}B_3+\frac{2}{d(d-1)}B_2,\quad A_6=B_6, \quad 
A_7=B_7+\frac{1}{d}B_6\cr
&&A_8=\frac{d}{d-2}B_8+\frac{1}{2d}B_6+\frac{2}{d-2}B_{14}-
\frac{2d}{(d-2)^2}B_{13}-\frac{2}{(d-1)(d-2)}B_{11}\cr
&&A_9=B_9+\frac{4}{d-2}B_7+\frac{2}{d(d-1)}B_{6},\quad A_{10}=B_{10}\cr
&&A_{11}=B_{11}+\frac{1}{d}B_{10},\quad 
A_{12}=B_{12}+\frac{4}{d-2}B_{11}+\frac{2}{d(d-1)}B_{10}\cr
&&A_{13}=B_{13}+\frac{3}{d}B_{11}+\frac{1}{d^2}B_{10}\cr
&&A_{14}=B_{14}-\frac{2}{d-2}B_{13}+\frac{2d-3}{d(d-1)}B_{11}
+\frac{1}{d^2}B_{10}\cr
&&A_{15}=B_{15}+\frac{4}{d-2}B_{14}+\frac{2(d-4)}{(d-2)^2}B_{13}
+\frac{1}{d}B_{12} +\frac{4(2d-3)}{d(d-1)(d-2)}B_{11}
+\frac{2}{d^2(d-1)}B_{10}\cr
&&A_{16}=B_{16}+\frac{12}{d-2}B_{15}+\frac{24}{(d-2)^2}B_{14}
+\frac{8(d-4)}{(d-2)^3}B_{13}\cr
&&\qquad +\frac{6}{d(d-1)}B_{12} +\frac{24}{d(d-1)(d-2)}B_{11}
+\frac{4}{d^2(d-1)^2}B_{10}
\cr
&&A_{17}=B_{17}-\frac{3}{d-2}B_{15}+\frac{3(d-4)}{(d-2)^2}B_{14}
+\frac{2(8-3d)}{(d-2)^3}B_{13}\cr
&&\qquad -\frac{3}{2d(d-1)}B_{12} +\frac{3(d-4)}{d(d-1)(d-2)}B_{11}
+\frac{d-2}{d^2(d-1)^2}B_{10}\ . 
\nonumber
\eea
One can show that the following linear combinations 
\bea
&&C_1=B_1,\quad C_2=B_2+B_6,\quad C_3=B_3+B_7,\quad 
C_4=B_4+B_8,\quad
C_5=B_5+B_9,\cr
&&C_6=\frac{(d-2)^2}{4d^2}B_2-B_4-\frac{1}{d-1}B_{11}-
\frac{d}{d-2}B_{13}+B_{14}\nonumber
\\
&&C_7=\frac{(d-2)(d-3)}{4d^2(d-1)}B_2-\frac{d-3}{d-2}(B_3-B_4)
+\frac14 B_5+\frac{1}{2d}B_{12}+
\frac{1}{2}B_{15}-\frac14 B_{16}-B_{17}\nonumber
\eea
are total derivatives. These are the important identities used in
the main text.

The basis of invariants $V_p$ 
depending on the curvature $F_{ij}$ and the 
endomorphism $E$ is 
\bea
&&V_1= \n_kF_{ij}\n^kF^{ij},\quad V_2=\n_jF_{ij}\n^kF^{ik},\quad 
V_3= F_{ij}\nabla^2 F^{ij},\quad V_4=F_{ij}F^{jk}F_{k}{}^i,\cr
&&V_5=  R_{mnij}F^{mn}F^{ij} , \quad
V_6= R_{jk}F^{jn}F^k{}_n,\quad
V_7= RF_{ij}F^{ij},\quad 
V_8= \nabla^4 E,\quad
V_9= E \nabla^2 E,\cr
&&V_{10}=\n_k E\n^k E,\quad
V_{11}= V^3 ,\quad
V_{12}= E F_{ij}^2 ,\quad
V_{13}= R \nabla^2 E ,\quad
V_{14}= R_{ij}\n^i\n^jE ,\cr
&&V_{15}= \n_kR\n^kE,\quad V_{16}= EER,\quad
V_{17}= E\nabla^2 R,\quad
V_{18}= ER^2,\cr
&& V_{19}=ER_{ij}^2,\quad 
V_{20}=ER_{ijkl}^2\ . 
\nonumber
\eea


\section*{Appendix B:\  Heat kernel expansion and $b_6$ coefficient
}

The heat kernel coefficients for a general Laplace 
operator of the form
$\D= -\nabla^2 - E$ 
with connection 
of curvature $F_{ab}$ as defined in appendix A
and matrix potential $E$,
 were
computed,  up to and including  $b_6$, 
  in \cite{Gil}.
For convenience, we present the explicit form 
of 
these leading terms in the  heat kernel  
expansion below 
  and use this opportunity
to cast  this expansion 
 into  a form that may be advantageous for certain computational 
purposes. In principle, one can compute  various terms
of the heat kernel expansion by using the standard 
perturbation theory for 
 a quantum mechanical path integral \cite{Ba1, Ba2}.
The latter naturally separates connected and disconnected 
particle theory 
diagrams and suggests  the following 
representation of the heat kernel expansion
$$
{\rm Tr} \biggl [ \s(x) e^{ -s \D} \biggr] =
{1\over (4\pi s)^{d\over 2}} 
{\rm Tr} \biggl [ \s(x) \sum_{n=0}^\infty a_{2n} s^n 
\biggr ]
={1\over (4\pi s)^{d\over2}} {\rm Tr} \biggl [ \s(x) 
\exp( \sum_{n=1}^\infty \alpha_{2n} s^n)
\biggr ]\ . 
$$
The standard $b_{2n}$ coefficients  are then 
$$b_{2n}= {1\over (4\pi)^{d\over 2}} a_{2n}
\ , \ \ \  \ a_{2n} =  \alpha_{2n} +  \beta_{2n}\ , $$
where $\alpha_{2n}$ and  $\beta_{2n}$
indicate the parts coming from  connected and disconnected 
quantum mechanical diagrams, respectively.
In the above expression 
 $\sigma(x)$ is an arbitrary function and 
${\rm Tr}(...) \equiv \int_M d^dx \sqrt{g}\, {\rm tr}_V(...)$.
Using the cyclicity of the trace one finds
$$
\beta_0 =0, \ \ \
\beta_2 =0, \ \ \
\beta_4 = {1\over 2} \alpha_2^2 , \ \ \
\beta_6 = {1\over 6} \alpha_2^3 + \alpha_2 \alpha_4\ . 
$$
Then  the formulas of \cite{Gil} for 
$b_0,b_2,b_4,b_6$ imply that 
\bea 
\a_0 &=& 1 \cr
\a_2 &=& E+{1\over 6} R \cr
\a_4 &=& {1\over 6} \nabla^2 \biggl ( E+{1\over 5} R \biggr )
+{1\over  180} (R_{abmn}^2 -R_{ab}^2) + {1\over 12} F_{ab}^2  \cr
\a_6 &=& {1\over 7!} \biggl [
18 \nabla^4 R + 17 (\nabla_a R)^2  -2 (\nabla_a R_{mn})^2
-4 \nabla_a  R_{bm} \nabla^b  R^{am} \cr
&& +9 (\nabla_a R_{mnij})^2 -8 R_{ab}\nabla^2 R^{ab}
+ 12 R_{ab} \nabla^a   \nabla^b R 
+12 R_{abmn}\nabla^2 R^{abmn}\cr
&& +{8\over 9} R_a{}^m R_m{}^i R_i{}^a 
+ {8\over 3} R_{ab} R_{mn} R^{amnb} 
- {16\over 3} R_{ab} R^a{}_{mnl} R^{bmnl}
\cr
&& + {44\over 9} R_{ab}{}^{mn} R_{mn}{}^{ij} R_{ij}{}^{ab}
-{80 \over 9} R_{iabj} R^{amnb} R_m{}^{ij}{}_n \biggr ] \cr
&& + {2\over 6!} \biggr[ 8 (\nabla_a F_{mn})^2 +2 (\nabla^a F_{am})^2
+12 F_{ab}\nabla^2 F^{ab} - 12 F_a{}^mF_m{}^i F_i{}^a \cr
&& + 6 R_{abmn} F^{ab} F^{mn} -4 R_{ab} F^{am} F^b{}_m
+ 6 \nabla^4 E  + 30 (\nabla_a E)^2 \cr
&&+ 4 R_{ab} \nabla^a\nabla^b E + 12 \nabla_a R \nabla^a E \biggr]
\ .
\nonumber \eea
In terms of the $A_s$ and $V_p$  invariants  of  Appendix A
the expression for 
$b_6 = { 1 \ov (4\pi)^3}
 (  \a_6 +{1\over 6} \alpha_2^3 + \alpha_2 \alpha_4)
  $ reads as in eq. (\ref{b6}).


\section*{Appendix C:\  Some $b_6$ coefficients in arbitrary $d$}

The coefficient $b_6^{(1)}$ of the Hodge-DeRham operator 
$\D^{(1)}$ acting 
on 1-forms in a $d$-dimensional manifold is given by
\bea
b_6^{(1)}
&=&\frac{1}{(4\pi )^{\frac{d}{2}}\, 7!}\biggl( 
(18d-84)A_{1}+(17d-168)A_{2}
+(364-2d)A_{3}+(56-4d)A_{4}\cr
&+&(9d-112)A_{5}+(28d-308)A_6
+(840-8d)A_{7}+(24d-112)A_{8}+(12d-168)A_{9}\cr
&+&(\frac{35d}{9}-70)A_{10}+(448-\frac{14d}{3})A_{11}+
(\frac{14d}{3}-98)A_{12}
+(-\frac{208d}{9}-728)A_{13}\cr
&+&(\frac{64d}{3}-112)A_{14}+(476-\frac{16d}{3})A_{15}
+(\frac{46d}{9}-84)A_{16}+(\frac{80d}{9}-168)A_{17}\biggr)\ . 
\nonumber
\eea
To compute the coefficient $b_6^{(2)}$ of the Hodge-DeRham 
operator $\D^{(2)}$ acting on 2-forms, we need several relations involving 
the curvature tensor of
the connection and the endomorphism 
\bea
\l F_{ij}\r_{ab}^{cd}&=&2R_{ij[a}{}^{[c}\d_{b]}^{d]}\ ,  \cr
E_{ab}^{cd}&=&-2 R_{[a}^{[c} \d_{b]}^{d]}
+ R_{ab}{}^{cd} \ . 
\nonumber
\eea
One can show that
\bea
{\rm tr}\,( F_{ij}F_{kl})&=&(2-d)R_{ijab}R_{klab}\cr
{\rm tr}\, (F_{ij}F_{kl}F_{mn})&=&(d-2)R_{ijab}R_{klbc}R_{ca}\cr
{\rm tr}\, E^2&=&R_{ijab}R_{ijab}+(d-6)R_{ab}R_{ab}+R^2\cr
{\rm tr}\, E^3&=&R_{ijab}R_{abkl}R_{klij}-6R_{iabc}R_{jabc}R_{ij}
-6R_{ijab}R_{ja}R_{ib}\cr
&+&(10-d)R_{ab}R_{bc}R_{ca}-3RR_{ab}R_{ab}\cr
{\rm tr}\,( EF_{ij}F_{ij})&=&R_{ijab}R_{abkl}R_{klij}
+(d-6)R_{iabc}R_{jabc}R_{ij}+RR_{abcd}R_{abcd}\ . \nonumber
\eea
Using these relations, we obtain $b_6^{(2)}$
for an arbitrary $d$-dimensional manifold
\bea
&&b_6^{(2)}
=\frac{1}{(4\pi )^{\frac{d}{2}}\, 7!}\biggl( (9d^2-93d+168)A_{1}
+(\frac{17d^2}{2}-\frac{353d}{2}+756)A_{2}\cr
&&+(-d^2+365d-2408)A_{3}+(-2d^2+58d-112)A_{4}
+(\frac{9d^2}{2}-\frac{233d}{2}+644)A_{5}\cr
&&+
(14d^2-322d+1456)A_6+(-4d^2+844d-5040)A_{7}
+(12d^2-124d+224)A_{8}\cr
&&+(6d^2-174d+1176)A_{9}+(\frac{35d^2}{18}- \frac{1295d}{18}+560)A_{10}
+(-\frac{7d^2}{3}+\frac{1351d}{3}-5096)A_{11}\cr
&&+(\frac{7d^2}{3}-\frac{301d}{3}+1036)A_{12}
+(\frac{104d^2}{9}-\frac{6448d}{9}+8176)A_{13}
+(\frac{32d^2}{3}-\frac{368d}{3}+5264)A_{14}\cr
&&+(-\frac{8d^2}{3}+\frac{1436d}{3}-7672)A_{15}
+(\frac{22d^2}{9}-\frac{778d}{9}+1428)A_{16}
+(\frac{40d^2}{9}-\frac{1552d}{9}+336)A_{17}\biggr)\ . 
\nonumber
\eea

\vskip 0.5cm
{\bf Acknowledgements}

F.B. would like to thank D. Anselmi for useful discussions.
A.A.T. is grateful to  M. Henningson, Yu. Obukhov and 
 K. Skenderis for helpful discussions and correspondence.
 We are also grateful to H. Osborn for an important comment
 on the first version of the paper.
The work of S.F. was supported by
the U.S. Department of Energy under grant No. DE-FG02-96ER40967.
The work of A.T. was  supported in part by
the DOE  grant No. DOE/ER/01545-783, 
 the EC TMR programme grant ERBFMRX-CT96-0045,  
INTAS grant No.96-538 and  NATO grant
 PST.CLG 974965.


\baselineskip=14pt

\end{document}